\documentclass[12pt]{article}
\usepackage{amsfonts}
\begin{document}
\begin{center}
{\normalsize DRAG AND LIFT FORCES FOR THE THIN BODY IN SUPERSONIC FLOW OF THE NONEQUILIBRUM GAS}\\
\medskip
V.V. Maximov, E.Ya. Kogan, I.P. Zavershinsky and  A.P.Zubarev\\
Samara  State Aerospace University, Moskow str. 34, Samara, 443086, Russia
\end{center}
\medskip
{\small The classical problem of the fluid mechanics is the problem of a supersonic motion
around a thin body was generalized to the case of non-equilibrium gas. The drag and lift
force coefficients were founded. It is shown that the drag and lift force coefficients in
the acoustically active supersonic flow are both decreased.}
 
\bigskip
{\normalsize \bf Introduction.}
 
The classical problem of the fluid mechanics is the problem of a supersonic motion  around a thin body. One of the problem demanding generalisation on a case of an non-equilibrium media, is the problem of a supersonic flow past the thin body, [1,2].
 
This problem, in particular, was investigated in [3,4] in linear approach for a case of inviscid flow of non-equilibrium gas. The unwidely expressions for lifting force which are not admitting generalisations on a non-linear case were received. The results of experimental and numerical investigations of the properties of supersonic flows around blunt bodies in the presence of electric discharge in stream and in the front of the body were given in [5,6]. The essential decreasing of the drag force was observed.
 
The present paper considers the supersonic motion of the non-equilibrium gas around a thin body
up to the order of $\sim \varepsilon^3$, where $\varepsilon$ is a small parameter.
Drag and lift force, as is known, connected with the impulse which is carried away by shock
waves. However in acoustically active media the structure of a acoustic field is defined
by second viscosity coefficient $\mu \gg \eta$, [7,8], where $\eta$  is the first (shear)
viscosity coefficient. Since the acoustic field structure around a body
is sufficiently changed. Therefore the neglect by dissipative processes becomes incorrect, [9].
 
We consider a plain supersonic flow ($M = v_0/u_s >1$ is the Mach number, $v_0$
is flow speed far from a body, $u_s$ is sound velocity) around a thin body ($l_x/l_y \gg M$)
with a small attack angle $\delta = l_y/l_x \ll 1$. Here $l_y$ is effective thickness, $l_x$
is length of a body. Further we examine the integrated characteristics of a flow - drag and
lift force coefficients $C_{x;y}$
\begin{eqnarray}
\nonumber
C_{x;y}=(1/2\rho_0 v_0^2)\int Pn_{x;y}ds
\end{eqnarray}
where $\rho_0$ is a gas density in infinity, P is a gas pressure in a point (x,y), $n_{x;y}$ are components of a normal vector. The integration is conducted along an airfoil surface. Pressure on an airfoil surface can be represented in form $P = P_0 + P_1$, where $P_0$ is a gas pressure in infinity, $P_1 = \rho_0 v_0v_1 - \rho_0(v_x^2 + v_y^2)/2$ is disturbance of pressure, $v{x;y}$ are components of the flow velocity perturbations.
 
\bigskip
{\normalsize \bf Statement of the problem and main results}
 
The assumption formulated permit to use of the equations which describe the fluid motion in form of Navier-Stoces which must be completed with the equation of state and energy. In the non-equilibrium gas for perturbations of parameters these equations up to the of (2 order for the high-frequency perturbations (( " 1 lead to the following equation for the velocity potential ( (v = (()
\begin{eqnarray}
\frac{\partial^2{\phi}}{\partial{y}^2}\;-\;(\,M^2-1\,)\frac{\partial^2{\phi}}{\partial x^2}\;-\;
\frac{M}{u_\infty}\left([2\,+\,(\gamma-1) M^2]
\frac{\partial \phi} {\partial x}\frac{\partial^2 \phi}{\partial x^2}\;+\;
2\frac{\partial \phi}{\partial y}
\frac{\partial^2 \phi}{\partial x \partial y}\right)\,- \\
\nonumber
-\,\frac{\mu_\infty M^3}{u_\infty}\frac{\partial^3 \phi}{\partial x^3}\,+\,\alpha_\infty M \frac{\partial \phi}{\partial x}\;=\;0.
\end{eqnarray}
The similar equations are well known in the theory of non-linear ion-sound waves, gravitational waves on a surface of a liquid etc. Here $\alpha_\infty$  is spatial increment of acoustic instability within the limits of frozen relaxation of internal degrees of freedom $\omega \tau \gg 1$, $\mu_\infty = 4 \eta/3 + \kappa(1/c_{V \infty} - 1/c_{P \infty})$. With a conclusion of Eq.(2) it was necessary, that the characteristic length of energy dissipation  $\alpha_\infty^{-1}$ is small in comparison with longitudinal scale of a body $l_x$. For the decision of the problem, the Eq.(1) needs to be added by a boundary condition for normal components of speed of gas. In case of firm walls
\begin{eqnarray}
\left(U\,+\,\frac{\partial \phi}{\partial x}\right)n_x\;+\;
\left.\frac{\partial \phi}{\partial y}n_y \right|_{y\rightarrow\pm0}\;=\;0
\end{eqnarray}
In space above a body (y > 0) it is possible to consider $n_x = - \partial \zeta_2/ \partial x$ and under a body $n_x = - \partial \zeta_1/ \partial x$, where $y = \zeta_{2;1}(x)$ are equations of the top and bottom profile surface.
 
For the further simplification of considered model the method of the consecutive approaches is used, by choosing as the initial decision of the Eq.(1), as a simple wave. In result we could have
\begin{eqnarray}
\frac{\partial u}{\partial y}\,\pm\, \beta _{\infty}\frac{\partial u}{\partial x}
\,\pm\, \frac{\Psi_{\infty} M^3}{2\beta_{\infty} u_{\infty}}u\frac{\partial u}{\partial x}
\;=\;\pm \mu_{\infty} \frac{M^3}{u_{\infty}}\frac{\partial^2 u}{\partial x^2}
\,\mp\,\frac{M\alpha_{\infty}}{u_\infty}u
\end{eqnarray}
Boundary condition will be transformed to a form
\begin{eqnarray}
u(y\,=\,\pm0)\;=\;\pm \frac{Mu_{\infty}}{\beta}\frac{\partial \zeta_{2;1}}{\partial x}\;\pm\; \mu_{\infty} \frac{M^4}{2u_{\infty} \beta^3} \frac{\partial^2 \zeta_{2;1}}{\partial x^2}\;\mp\; \frac{\alpha_\infty M^2 u_{\infty}}{2\beta^3}\zeta_{2;1}
\end{eqnarray}
The non-linear acoustic theory for non-equilibrium gas [8], predicts an opportunity of
stabilisation of acoustic instability developing with $\alpha_\infty < 0$, if $\mu_\infty > 0$,
in the second order of the theory of perturbations. The Eq. (3) has the limited decisions in
form of the sawtooth wave [8]. Other kind of the limited decisions with $\alpha_\infty < 0$,
if $\mu_\infty > 0$ this equation does not admit.
 
Passing to account of coefficients $C_{x;y}$, notice, that in a considered problem, by virtue of small $n_{x;y}$, expression for $P_1$  will be transformed to a kind: $P_1 = -\rho_0 v_0 \partial \phi/\partial x$, [1]. These coefficients are expressed through coefficient of pressure $C_P = (1/2\rho_0 v_0^2)P_1$ on a airfoil surfaces. The value $C_P$ on both surfaces up to the accepted accuracy is defined by expression:
\begin{eqnarray}
\nonumber
C_{p2;1}\;=\;\frac{2}{\beta }\left[\mp\frac{\partial \zeta_{2;1}}
{\partial x}\,\pm\,\frac{M \alpha_\infty}{\beta^2}\zeta_{2;1}\right]
\end{eqnarray}
 
Let's define angles of attack of the top and bottom lines of a profile to it chorde by expressions $\theta_{2;1}(x)= \partial \zeta_{2;1}/ \partial x  - \delta$. Let's notice, that here $\zeta_{2;1} = l_x, \zeta_{2;1}(l_x) = 0$. Then we receive
\begin{eqnarray}
C_x\;=\;\frac{2}{\beta}\left[2\delta^2\,+\,<\theta_1^2>_x\,+\, <\theta_2^2>_x \right], \;\;
C_y\;=\;\frac{2}{\beta_{\infty}} \left[2\delta\;+\; \frac{M \delta}{\beta^2}\alpha_{\infty} l_x \right].
\end{eqnarray}
Let's notice, that by replace of a  variables $\xi = ((\beta y-x)/l_x, z = y\Psi_\infty M^4 l_y/\beta^2l_x^2, \Re = u\beta l_x /Mu_\infty l_y$  Eq.(3) can be reduced to form
\begin{eqnarray}
\frac{\partial \Re}{\partial z}\,+\,\Re \frac{\partial \Re}{\partial \xi}\,=\,
\mu_{1\infty}\frac{\partial^2 \Re}{\partial \xi^2}\,-\,\alpha_{1\infty} \Re
\end{eqnarray}
where $\mu_{1\infty}\,=\,\mu_{\infty} \beta/\Psi_{\infty}M l_y u_{\infty}$, $\alpha_{1\infty}\,=\,\alpha_{\infty}\beta l_x^2/\Psi_{\infty}M^3 l_y u_{\infty}$.
This equation is similar under the form with received in [8] for acoustic perturbations in the motionless non-equilibrium media.
   The Eq. (3) and model, presented in [8] were received by means of the perturbation methods, use the same small parameters. Since the description of the acoustic field structure around the body up to the third order of the perturbation theory can be received by means of the variables transformations in the basic equations of the same order, presented in [8]. After transformations, in approach $\omega \tau\,\gg\, 1$ we come to the following results:
\begin{eqnarray}
\nonumber
\frac{\partial u}{\partial y}\pm\beta \frac{\partial u}{\partial x}
\pm \frac{\Psi_{\infty} M^3}{\beta_\infty u_{\infty}}u\frac{\partial u}{\partial x}
\mp k_{3\infty} \frac{M^2}{\beta u_{\infty}^3}\frac{l_x}{l_y}\frac{\partial u^3}{\partial x}
\,=\,\pm \frac{M B_{\infty}}{2u_{\infty}^2}\int udx
\pm \mu_{\infty} \frac{M^3}{u_{\infty}}\frac{\partial^2 u}{\partial x^2}\pm\\
\pm k_{1\infty} \frac{M^3}{\beta u_{\infty}^2}u\frac{\partial^2 u}{\partial x^2}
\pm k_{2\infty} \frac{M_{\infty}^3}{\beta u_{\infty}^2}\frac{\partial^2 u^2}{\partial x^2}
\mp \frac{M_{\infty}^3 \Psi_{\infty} \alpha_{\infty}}{\beta u_{\infty}} \frac{\partial u}{\partial x}\int udx \mp\frac{M \alpha_\infty}{\beta u_\infty} u
\pm \frac{M^3 \nu_{\infty }}{\beta u_{\infty}} u^2
\end{eqnarray}
\begin{eqnarray}
\nonumber
u(y=\pm0)\,=\,\pm \frac{M u_{\infty}}{\beta} \frac{\partial \zeta_{2;1}}{\partial x}+
\mu_{\infty} \frac{M^3}{\beta^2 u_{\infty }}\frac{\partial u}{\partial x}\,-
\,\frac{\Psi_{\infty} M^3}{\beta^2 u_{\infty}}u^2+
\frac{M B_{\infty }}{2u_{\infty}}\int(\int udx)dx+\\
\nonumber
+k_{1\infty} \frac{M^3}{\beta^2 u_{\infty}^2}\int u\frac{\partial^2 u}{\partial x^2}dx+
k_{2\infty} \frac{M^3}{\beta^2 u_{\infty}^2}\frac{\partial u^2}{\partial x}+
k_{3\infty}\frac{M^2}{\beta_{\infty}^2 u_{\infty}^3}\frac{l_x}{l_y} u^3-\\
\nonumber
-\frac{M^3 \Psi_{\infty} \alpha_{\infty}}{\beta^2 u_{\infty}}
\int (\frac{\partial u}{\partial x} \int udx) dx-
\frac{M \alpha_{\infty}}{\beta^2} \int udx +
\frac{M^3 \nu_{\infty}}{\beta^2 u_{\infty}} \int u^2 dx
\end{eqnarray}
 
Let's consider, for simplicity, a plain profile $\theta_{2;1}= 0$. The calculations completely similar mentioned above, allow to receive expressions for drag and lift force coefficients:
\begin{eqnarray}
\nonumber
C_x \,=\,\frac{2}{\beta} \delta^2 \left[2\,+\, \frac{M}{2 \beta}\alpha_{\infty} l_x \,+\,
\delta \frac{ k_{3\infty}M^3}{\beta u_{\infty}}\right],
\end{eqnarray}
\begin{eqnarray}
C_y \,=\,\frac{2}{\beta} \delta \left[2\,+\,\frac{M}{2 \beta}\alpha_{\infty} l_x-
\frac{M}{6 \beta u_{\infty}}B_{\infty} l_x^2 \,-\,
\delta \frac{ k_{3\infty}M^3}{\beta u_{\infty}}\right]
\end{eqnarray}
 
\bigskip
{\normalsize \bf Conclusions.}
 
   It is shown that on the base of the results of non-linear theory of the acoustic waves in the non-equilibrium media [8] the non-linear structure of the acoustic field, generated by profile in the supersonic flow of non-equilibrium gas up to the third order of the perturbations theory can be described easily.
The integrated characteristics of a flow around the airfoil (drag and lift force coefficients $C_{x;y}$) in the non-equilibrium gas were received.
 
The presented results demonstrate dependence of the drag and lift forces in the non-equilibrium gases from the degree of non-equilibrium. The drag and lift force coefficients in the acoustically active supersonic flow are both decreased. The given effect is connected with the appearance of the additional pressure gradient $\nabla P_{noneq} \sim \alpha_\infty l_x$, directed opposite to the classical gradient $\nabla P_{eq}$.
 
\bigskip
{\normalsize \bf Acknowlegements.}
 
This work is supported by RFBR grant N 96-02-16301.
\pagebreak
\begin{center}
{\raggedright \bf References}
\end{center}
\medskip
\begin{description}
\item 1. L.D. Landau., E.M. Lifshits, Theoretical physics (in Russian), 6, Nauka (1988 ).
\item 2. V.I. Karpman, Non-linear waves in dispersive media (in Russian), Nauka (1973).
\item 3. W. Vincenti, J. Fluid. Mech., 6, 481-496 (1959).
\item 4. D. Homentchvschi, ZAMM, 57, 461-469, (1977).
\item 5. A.U. Gridin, B.G. Efimov, A.V  Zabrodin et al, G.Keldysh's Institute of Applied Mathematics Prepr. N16 (in Russian). 1- 31 (1995).
\item 6. V.L.  Bychkov et al. G.Keldysh's Institute of Applied Mathematics Prepr. N27 (in Russian), 1-50 (1997).
\item 7. E.Ya. Kogan, N.E. Molevich, Izvestiya Vuzov. Fizica. (in Russian), 29, 53-58 (1986).
\item 8. I.P. Zavershinsky, E.Ya. Kogan, N.E Molevich,  Izvestiya Vuzov, Appl. Non-linear. Dyn., (in Russian) 3-4, 87-96, (1993).
\item 9. E.Ya. Kogan, N.E. Molevich, Acoustical physics, 36,  1431-1435 (1995).
\end{description}
\medskip
\end{document}